\documentclass[a4paper, reprint, nofootinbib, showkeys]{revtex4-2}

\usepackage{geometry}
\usepackage{graphicx}
\usepackage{float}

\usepackage[x11names]{xcolor}
\usepackage{enumitem}
\usepackage{lipsum}

\usepackage[utf8]{inputenc}
\usepackage[english]{babel}

\usepackage[acronym]{glossaries}
    \makeglossaries
    \newacronym{hnn}{HNN}{Hopfield Neural Network}
    \newacronym{ann}{ANN}{Artificial Neural Networks}
    \newacronym{rnn}{RNN}{Recurrent Neural Network}
    \newacronym{am}{AM}{Associative Memory}
    \newacronym{dam}{DAM}{Dense Associative Memory}
    \newacronym{sna}{SNA}{Signal--to--Noise Analysis}

\usepackage{mathtools}
\usepackage{amsthm, amssymb, bm}
\usepackage{wasysym}
    \let\vec\bm
    \DeclareMathOperator{\sign}{sign}
    \DeclareMathOperator{\ord}{\mathcal{O}}

    \DeclareMathOperator{\var}{Var}
    \DeclareMathOperator{\me}{e}
    
    \DeclarePairedDelimiter\ton{(}{)}
    \DeclarePairedDelimiter\qua{[}{]}
    \DeclarePairedDelimiter\gra{\{}{\}}
    \DeclarePairedDelimiter\abs{\|}{\|}
    \DeclarePairedDelimiter\mean{\langle}{\rangle}

\usepackage[pdftex, pdftitle={Article}, pdfauthor={Author}]{hyperref}
    \hypersetup{colorlinks=true, 
        citecolor=black, 
        linkcolor=black, 
        urlcolor=blue}

\makeatletter
    \renewcommand{\eqref}[1]{Eq.~\textup{\tagform@{\ref{#1}}}}
    
\makeatother

\begin{document}

    \title{Effect of spatial correlations on Hopfield Neural Network and Dense Associative Memories}
    \author{Giordano De Marzo$^{1,2,3,4}$}
    \email{giordano.demarzo@cref.it}
    \author{Giulio Iannelli$^{1,5}$}
    \email{giulio.iannelli@cref.it}
    \affiliation{$^1$Centro Ricerche Enrico Fermi, Piazza del Viminale, 1, I-00184 Rome, Italy.}
    \affiliation{$^2$Dipartimento di Fisica Universit\`a ``Sapienza”, P.le A. Moro, 2, I-00185 Rome, Italy.}
    \affiliation{$^3$Sapienza School for Advanced Studies, ``Sapienza'', P.le A. Moro, 2, I-00185 Rome, Italy.}
    \affiliation{$^4$Complexity Science Hub Vienna, Josefstaedter Strasse 39, 1080, Vienna, Austria.}
    \affiliation{$^5$Dipartimento di Fisica, Università di Roma “Tor Vergata”, 00133 Roma, Italy.}
    \date{\today} 
\begin{abstract}
    Hopfield model is one of the few neural networks for which analytical results can be obtained. However, most of them are derived under the assumption of random uncorrelated patterns, while in real life applications data to be stored show non-trivial correlations. In the present paper we study how the retrieval capability of the Hopfield network at null temperature is affected by spatial correlations in the data we feed to it. In particular, we use as patterns to be stored the configurations of a linear Ising model at inverse temperature $\beta$. Exploiting the signal to noise technique we obtain a phase diagram in the load of the Hopfield network and the Ising temperature where a fuzzy phase and a retrieval region can be observed. Remarkably, as the spatial correlation inside patterns is increased, the critical load of the Hopfield network diminishes, a result also confirmed by numerical simulations. The analysis is then generalized to Dense Associative Memories with arbitrary odd-body interactions, for which we obtain analogous results. 
\end{abstract}

\maketitle
    \section{Introduction}
The \gls{hnn} \cite{hopfield1982neural} stands as a cornerstone in the history of \gls{ann} and the scientific attempt of modeling brain functions \cite{amit1992modeling, coolen2005theory}. In particular the \gls{hnn} is a recurrent neural network composed of $N$ neurons or spins which interact pairwise and works as an \gls{am}, meaning that is it can store and then retrieve information.\par

A central point in the study of the \gls{hnn} is understanding the maximal amount of data this system can successfully memorize \cite{amit1987statistical}. It has been shown that an \gls{hnn} with $N$ neurons can store $K=\ord(N)$ i.i.d. random patterns, each composed of $N$ i.i.d. Rademacher variables, if some extent of faults is tolerated \cite{amit1985storing}. The storage capacity $\alpha=K/N$ of the \gls{hnn} can thus be of order one and the neural network still keeps working. More precisely it has been shown \cite{amit1985storing} that retrieval is possible up to $\alpha\leq\alpha_c$ with $\alpha_c\approx0.14$. Heuristically this can be understood noticing that the interaction matrix is made of $O(N^2)$ independent components and so the maximal amount of bits which can be stored must be of the same order of magnitude.\par

In order to increase the storage capacity of such system some adaptations of the \gls{hnn} has been proposed (e.g. \cite{folli2017maximum, storkey1997increasing}). In recent years \cite{krotov2016dense, krotov2020large} much attention has been devoted to generalized Hopfield networks with $p$-body interactions; this type of neural networks are called \glspl{dam}. The interaction matrix of a \gls{dam} with $p$-body interactions contains $O(N^p)$ components, so that, in principle, such network can store up to $N^{p-1}$ i.i.d. Rademacher patterns \cite{baldi1987number, venkatesh1991programmed}. These architectures have been proven not only to have a larger storage capacity, but also to be very robust against noise, faults and adversarial examples \cite{agliari2020neural, agliari2020tolerance, krotov2018dense}.\par

However patterns with random uncorrelated components, while representing an appealing standard for analytical studies, do not properly embody \emph{realistic} memories. This point is very relevant since \gls{hnn} and \glspl{dam} has been shown \cite{barra2012equivalence, tubiana2017emergence, mezard2017mean, barra2018phase, marullo2021boltzmann} to be formally equivalent to Boltzmann machines \cite{ackley1985learning, hinton2002training}, which are used in a number of machine learning applications where data are definitely not i.i.d. Rademacher patterns. Indeed in most real life situations patterns can be correlated one with the other or they may be structured, thus showing non trivial spatial correlations \cite{lowe1998storage, fontanari1990generalization, agliari2022emergence}. This last situation occurs, for instance, if the patterns to be stored are texts, which can be represented as binary uni-dimensional sequences whose entries are correlated due to grammatical and semantic rules. Analogously black and white images are binary matrices with spatial correlations since a black (or white) pixel is more likely to be surrounded by pixels of the same kind. However, despite the importance of patterns with such kind of spatial correlations, very few studies considered how \gls{hnn} behaves in this situation. In \cite{lowe1998storage} it has been shown that, if the spatial correlations arise from a simple two states Markov chain, than the \gls{hnn} still exhibits a critical capacity of order one. More realistic correlations, arising from a two-dimensional Ising model, have been studied in \cite{lowe2005storage}, but for the case of exact retrieval of the patters, thus providing only a lower bound to the true storage capacity. The effects of correlations on \glspl{dam} have been considered recently \cite{upgang2020storage} by using patterns generated by a Curie-Weiss model, but again only for a perfect retrieval of the patterns.\par

In the present work we considered how the \gls{hnn} behaves in presence of patterns generated by a one dimensional Ising model (or \emph{Ising linear chain}) at inverse temperature $\beta$. The spin configurations of this model can be regarded as prototypes of textual data. Note that by varying the temperature of the Ising model we can control the level of internal correlation of the patterns, while we set the temperature of the Hopfield model to zero, thus neglecting the effects of fast noise. In the rest of the work we refer to $\beta$ as the temperature parameter relative to the Ising model.\par

By using a \gls{sna} \cite{amit1992modeling} we asses the maximal storage capacity of the \gls{hnn} as function of $\beta$, finding that for $\beta>0$ the neural network can still store a number of patterns which grows linearly with the number of neurons $N$, even if the storage capacity is lower than in the case of i.i.d. Rademacher patterns. This result is non trivial since naively one would expect the storage capacity to increase due to the fact that internal correlations diminish the amount of information to be stored.  Moreover, differently from previous studies, we obtain the true maximal storage capacity and not a lower bound. This allows to draw a phase diagrams in the space defined by the parameters $(\alpha, \beta)$, where a retrieval region and a fuzzy phase can be identified. The analysis is then generalized to the \gls{dam} version of the \gls{hnn} with odd $p$--bodies interaction. We show that the maximal storage capacity deceases as $\beta$ increases, but again the network can memorize $O(N^{p-1})$ patterns as in the uncorrelated case, provided that $\beta>0$. \par

The paper is structured as follows:
    \begin{itemize}
        \item In Sec.~\ref{sec:s2n} we introduce the signal to noise technique and we apply it to the \gls{hnn} and to the \gls{dam} with i.i.d. Rademacher patterns. 
        \item In Sec.~\ref{sec:hopfield} we apply the \gls{sna} to the \gls{hnn} feed with patterns generated by a one dimensional Ising model and we compare the analytical results with numerical simulations.
        \item In Sec.~\ref{sec:dam} we generalize the results of the previous section to the case of a \gls{dam} with $p$--bodies interaction.
        \item Finally Sec.~\ref{sec:conclusions} is devoted to our conclusions and remarks.
    \end{itemize}
    \section{Signal to noise technique for uncorrelated patterns}
        \label{sec:s2n}
The \acrlong{hnn} is made of $N$ neurons associated with binary state variables, i.e. $\sigma_i=\pm 1$ for $i=1\cdots N$, and interacting pairwise on a complete graph with coupling matrix $J_{ij}$. Denoting by $\vec{\xi}^{\mu}$ with $\xi_i^{\mu}=\pm 1$ the $K$ binary patterns we want to store in the neural network, the coupling matrix is defined by the so called Hebbian rule, which reads
    \begin{flalign}
        \forall\,i\neq j && J_{ij} = \frac{1}{N}\sum_{\mu=1}^{K}\xi_i^{\mu}\xi_j^{\mu} &&
    \end{flalign}
and $J_{ii} = 0$ for all the diagonal elements, avoiding self interactions. The Hamiltonian of the system thus satisfies
    \begin{equation}
        H[\vec{\sigma}]=-\sum_{ij}^NJ_{ij}\sigma_i\sigma_j=-\frac{1}{N}\sum_{\mu=1}^{K}\sum_{i,j\neq i}^N\xi_i^{\mu}\xi_j^{\mu}\sigma_i\sigma_j.
        \label{eq:hamiltonian_hopfield}
    \end{equation}
In the following we will often avoid to explicitly write $i\neq j$, bearing in mind that the neurons have no self--interaction. If one neglects fast noise by considering the model at null temperature the evolution of the system is governed by the following dynamical rule
    \begin{equation}
        \sigma_i(t+1)=\sign\qua*{h_i\ton*{\vec{\sigma}(t)}},
            \label{eq:evolution}
    \end{equation}
where $h_i\ton*{\vec{\sigma}(t)}$ is the local field acting on the $i$th neuron at time $t$ and satisfying 
    \begin{equation}
        h_i\ton*{\vec{\sigma}(t)}=\frac{1}{N}\sum_{\mu=1}^{K}\sum_{j\neq i}^N\xi_i^{\mu}\xi_j^{\mu}\sigma_j(t).
            \label{eq:field}
    \end{equation}
Note that the Hamiltonian \eqref{eq:hamiltonian_hopfield} can be rewritten in terms of these local fields as $H[\vec{\sigma}]=-\sum_{i}h_i\ton*{\vec{\sigma}}\sigma_i$ and the dynamics defined by \eqref{eq:evolution} is equivalent to a steepest descent algorithm whose Lyapunov function is the Hamiltonian. Moreover from \eqref{eq:evolution} it follows that the state of a given spin $\sigma_i$ is unaltered by the dynamics if it holds $\sign\qua{\sigma_i}=\sign\qua{h_i\ton{\vec{\sigma}}}$, that is if it holds $\sigma_ih_i\ton{\vec{\sigma}}>0$. As a consequence a pattern, let us say $\vec{\xi}^{1}$, is a fixed point of the dynamics and is thus dynamically stable provided that its components satisfy 
    \begin{equation}
        \xi^{1}_ih_i\ton*{\vec{\xi}^{1}}=\frac{1}{N}\sum_{\mu=1}^{K}\sum_{j>1}^N\xi_1^{\mu}\xi_j^{\mu}\xi_j^{1}\xi_1^{1}>0.
        \label{eq:stability_patterns}
    \end{equation}
Here and in the following we focus, without loss of generality, on the first pattern. However this quantity, being a sum of random variables, is itself a random variable and so, in order to assess if it is positive, we have to determine whether its expectation is positive and larger than its standard deviation. The first quantity is the signal, while the latter is the noise affecting it. This explaining where the name ``signal to noise'' steams from. We denote by $\mean{\cdot}$ the expectation on the probability distribution of the $\xi_i^{\bar\mu}$ for fixed $\bar\mu$ and by $\var\qua{\cdot}$ the variance computed over the same distribution. In these terms patterns are dynamically stable if it holds
    \begin{equation}
        \sqrt{C}\mean*{\xi^{1}_ih_i\ton*{\vec{\xi}^{1}}}\geq\sqrt{\var\qua*{\xi^{1}_ih_i\ton*{\vec{\xi}^{1}}}},
        \label{eq:stability_patterns_mean_var}
    \end{equation}
where $C$ is a numerical constant fixing the level of tolerated faults. Note that by explicitly writing the variance this condition is equivalent to 
    \begin{equation}
        (C+1)\mean*{\xi^{1}_ih_i\ton*{\vec{\xi}^{1}}}^2\geq\mean*{\ton*{\xi^{1}_1h_1\ton*{\vec{\xi}^{1}}}^2}.
        \label{eq:stability_patterns_simpler}
    \end{equation}
At this point we have all the ingredient for applying the \acrlong{sna} technique to the \gls{hnn} fed with i.i.d. Rademacher patters. First we compute the mean value appearing in \eqref{eq:stability_patterns_mean_var}
    \begin{multline}
        \mean*{\xi^{1}_ih_i\ton*{\vec{\xi}^{1}}}=\frac{1}{N}\sum_{\mu=1}^{K}\sum_{j>1}^N\mean*{\xi_1^{\mu}\xi_j^{\mu}\xi_j^{1}\xi_1^{1}}=\\
        =\frac{N-1}{N}+\frac{1}{N}\sum_{\mu=2}^{K}\sum_{j>1}^N\mean*{\xi_1^{\mu}}\mean*{\xi_j^{\mu}}\mean*{\xi_j^{1}}\mean*{\xi_1^{1}} = 1,
        \label{eq:mean_hopfield_uncorrelated}
    \end{multline}
where we implicitly took the limit of large $N$ and we used the fact that in the case under consideration, since there are no correlations, the expectation factorizes. Moreover, being $P\ton*{\xi_i^{\mu}=+1}=P\ton*{\xi_i^{\mu}=-1}=1/2$, it holds $\mean*{\xi_i^{\mu}}=0$. Then, in order to obtain the variance, we first compute the expectation of $\ton*{\xi^{1}_1h_1\ton*{\vec{\xi}^{1}}}^2$. By noticing that 
    \begin{multline}
        \qua*{\xi^{1}_1h_1\ton*{\vec{\xi}^{1}}}^2=\ton*{\frac{N-1}{N}+\frac{1}{N}\sum_{\mu=2}^{K}\sum_{j>1}^N\xi_1^{\mu}\xi_j^{\mu}\xi_j^{1}\xi_1^{1}}^2=\\
        =2h_1\xi_1^1-1+\frac{1}{N^2}\sum_{j, k>1}^N\sum_{\mu, \rho>1}^K\xi_1^{\mu}\xi_j^{\mu}\xi_j^{1}\xi_1^{\rho}\xi_k^{\rho}\xi_k^{1}
            \label{eq:expression_squared_field}
    \end{multline}
and by using \eqref{eq:mean_hopfield_uncorrelated} we get
    \begin{align*}
        \mean*{\ton*{\xi^{1}_1h_1\ton*{\vec{\xi}^{1}}}^2}&=1+\frac{1}{N^2}\sum_{j, k>1}^N\sum_{\mu, \rho>1}^K\mean*{\xi_1^{\mu}\xi_j^{\mu}\xi_j^{1}\xi_1^{\rho}\xi_k^{\rho}\xi_k^{1}}\\
        &=1+\frac{\ton*{K-1}\ton*{N-1}}{N^2}=1+\frac{K-1}{N}.
    \end{align*}
Here we used the fact that the only terms surviving the expectation are those for which $j=k$ and $\mu=\rho$. Combining this expression with \eqref{eq:mean_hopfield_uncorrelated} we can finally write the variance
    \begin{align}
        \var\qua*{\xi^{1}_ih_i\ton*{\vec{\xi}^{1}}}&=\mean*{\ton*{\xi^{1}_1h_1\ton*{\vec{\xi}^{1}}}^2}-\mean*{\xi^{1}_ih_i\ton*{\vec{\xi}^{1}}}^2 =\nonumber\\
        &=\frac{K-1}{N}.
        \label{eq:variance_hopfield_uncorrelated}
    \end{align}
Plugging this relation and \eqref{eq:mean_hopfield_uncorrelated} into the condition for the dynamical stability \eqref{eq:stability_patterns_mean_var} we thus obtain that the patterns are dynamically stable provided that $C\geq{K-1}/{N}$ that is if
    \begin{equation}
        \alpha\leq C.
            \label{eq:dynstabalpha_C}
    \end{equation}
Here we introduced the load $\alpha=K/N$ and we considered that, in the thermodynamic limit, $K\gg1$. Since $C$ is a numerical constant we thus see that the network can successfully store up to $K=O(N)$ patterns before they stop to be fixed points of the dynamics, recovering, as expected, what previously proven \cite{amit1985storing, amit1992modeling}.\par
In the vary same way we can analyze also the corresponding \gls{dam} with $p$--bodies interaction, whose Hamiltonian, in analogy with \eqref{eq:hamiltonian_hopfield}, is 
    \begin{align}
		H^{(p)}[\vec{\sigma}] &= -\sum_{i_1\ldots i_p}^NJ_{i_1\ldots i_p}\sigma_{i_1}\ldots\sigma_{i_p}=\nonumber\\ 
		&=-\frac{1}{N^{p-1}}\sum_{\mu=1}^K\sum_{i_1\ldots i_p}^N\prod_{k=i_1\ldots i_p}\xi_k^{\mu}\sigma_k.
		\label{eq:hamiltoniana_p_body}
	\end{align}
We can then define the field $h_i^{(p)}\ton*{\vec{\sigma}}$ acting on the $i$th spin at time $t$ as
    \begin{equation}
        h_i^{(p)}\ton*{\vec{\sigma}(t)}=-\frac{1}{N^{p-1}}\sum_{\mu=1}^K\sum_{i_2\ldots i_p}^N\xi_i^{\mu}\prod_{k=i_2\ldots i_p}\xi_k^{\mu}\sigma_k(t)
        \label{eq:local_field_pbody}
    \end{equation}
and in these terms the dynamical rule determining the evolution on the neural network as the same form as \eqref{eq:evolution}, that is
    \[
        \sigma_i(t+1)=\sign\qua*{h_i^{(p)}\ton*{\vec{\sigma}(t)}}.
    \]
As a consequence all we have to do is to repeat the computation we did in the case of the \gls{hnn} and use \eqref{eq:stability_patterns_mean_var} to asses the dynamical stability of the patterns. The only difference is that this time the local field $h_i^{(p)}$ is given by \eqref{eq:local_field_pbody} instead of \eqref{eq:field}. Focusing on $\xi_1^1$ and doing manipulations analogous to those already done for the \gls{hnn} (corresponding to $p=2$), we obtain
    \begin{equation}
        \mean*{\xi^{1}_ih_i^{(p)}\ton*{\vec{\xi}^{1}}} = 1,
        \label{eq:mean_pbody_uncorrelated}
    \end{equation}
and
    \begin{equation}
        \var\qua*{\xi^{1}_ih_i^{(p)}\ton*{\vec{\xi}^{1}}} = (p-1)!\frac{K-1}{N^{p-1}}.
        \label{eq:variance_pbody_uncorrelated}
    \end{equation}
Putting these equations into the condition for the dynamical stability we get $C\geq (p-1)!{K-1}/(N^{p-1})$, which means that \eqref{eq:dynstabalpha_C} rewrites as
    \begin{equation}
        \alpha^{(p)}\geq \frac{C}{(p-1)!}
    \end{equation}
where we defined the load of a $p$--bodies \gls{dam} as 
    \begin{equation}
        \alpha^{(p)}=K/N^{p-1}.
    \end{equation}
We thus see that in this case, as mentioned above and already proven, the neural network can store up to $O(N^{p-1})$ patterns.
    \section{Hopfield Neural Network with correlated patterns} 
        \label{sec:hopfield}
\begin{figure*}[!ht]
    \centering
    \includegraphics[width=\textwidth]{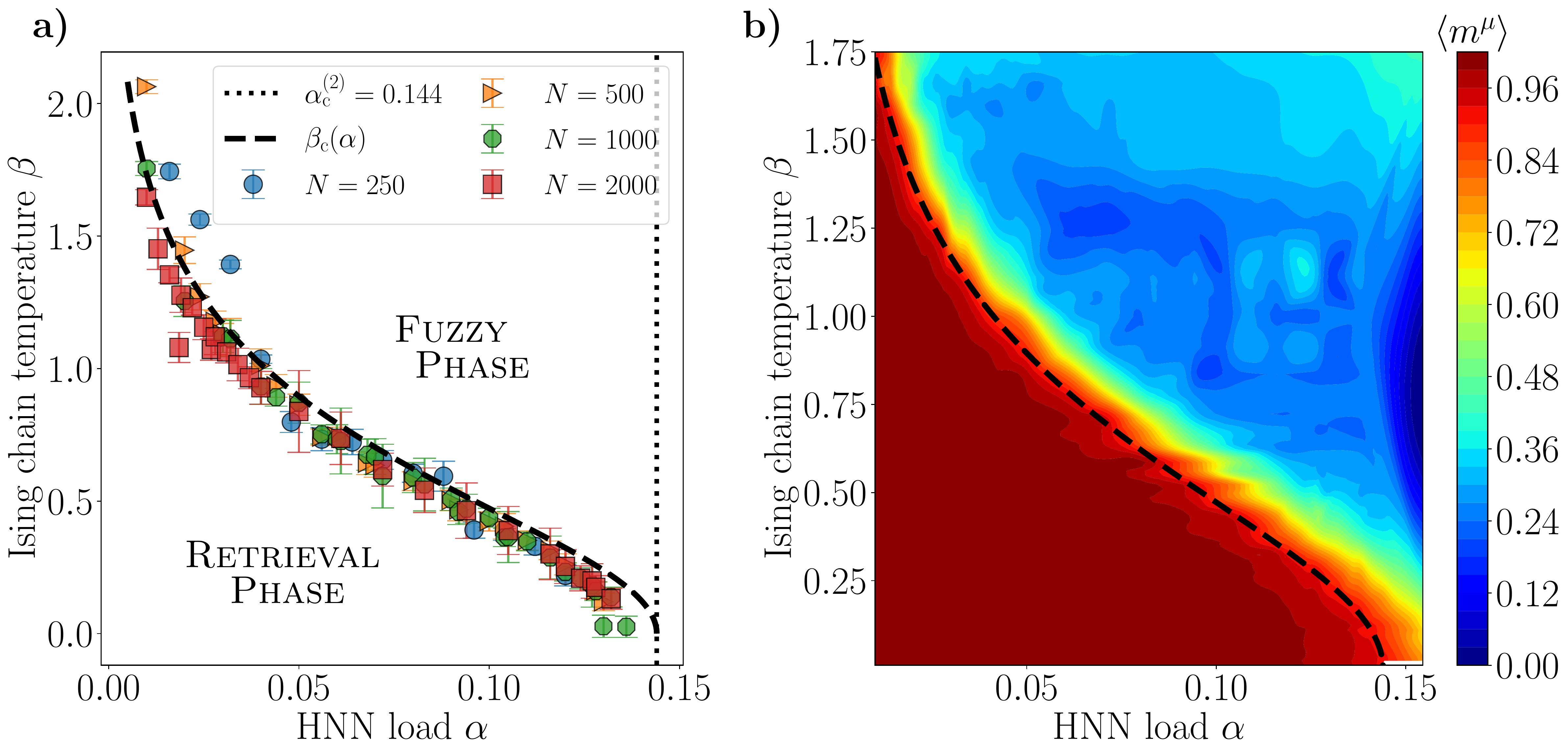}
        \caption{\textbf{Numerical simulations for the Hopfield network endowed with spatially correlated patters.} \textbf{a)} We simulated a Hopfield neural network with $N=250, 500, 1000, 2000$ neurons and varying load $\alpha$ under the dynamics \eqref{eq:evolution} and using as starting state one of the patterns $\vec{\sigma}=\vec{\xi^{\mu}}$ generated using a linear Ising chain at temperature $\beta$. We then collected the Mattis magnetization $m_{\mu}$ for each possible starting pattern, we repeated the process $M=5$ times and we computed the average magnetization $\mean*{m^{\mu}}$. For any $\alpha$ we computed the critical $\beta_c$ above which retrieval breaks as the value for which the average magnetization is equal to Sompolinsky's magnetization $m=0.967$ \cite{amit1985storing}. We also report the theoretical prediction as a solid black line. Uncertainties have been computed as the semi-difference between the last point above Sompolinsky's threshold and the one just below. \textbf{b)} Cubic interpolation of the Level curves of the magnetization for the same numerical simulations described above ($N=2000$). The theoretical curve separating retrieval from the fuzzy region is displayed as a dashed line. }\label{fig:H2-T_ic_vs_alphafull}
\end{figure*}
In the previous section we shown how \gls{hnn} behaves when fed with uncorrelated Rademacher patterns, but as aforementioned real data are structured and characterized by non trivial spatial correlations. Examples are text and images, both playing a central role in a huge number of machine learning applications.\par



It is then very natural to asses what happens to the \gls{hnn}, and so also to the formally equivalent Boltzmann machine, when the patterns have internal correlations (but different patterns are uncorrelated). In order to do so, we exploit the one dimensional Ising model, that we use to generate one dimensional correlated data which can be regarded as prototypes of texts. More precisely, each pattern is obtained as a configuration of the Ising model at inverse temperature $\beta$. By tuning $\beta$ we can then easily control the level of correlation inside the data. This means that now the probability distribution over which we make the expectation is no more the trivial Rademacher distribution $P\ton*{\xi_i^{\mu}=+1}=P\ton*{\xi_i^{\mu}=-1}=1/2$, but it is instead the Boltzmann one. More precisely it holds 
	\[
	    P\ton*{\vec{\xi}^{\mu}}=\frac{1}{Z}\me^{-\beta H_{I}\qua*{\vec{\xi}^{\mu}}}
	\]
	where $Z$ is the partition function and $H_{I}\qua*{\vec{\xi}^{\mu}}$ is the one dimensional Ising Hamiltonian
	\[
	     H_{I}\qua*{\vec{\xi}^{\mu}}=-\frac{1}{2}\sum_{i, j\neq i}\xi_i^{\mu}\xi_j^{\mu}.
	\]
	As a consequence, when taking the expectations, we now have to consider also correlations since the probability distribution does not factorizes. However, it is easy to show, for instance by using the transfer matrix formalism \cite{kramers1941statistics}, that in the 1D Ising model only the two points correlator is non-trivial, while all correlations involving an odd number of points are null
    \begin{equation}
		\mean*{\xi_i^{\mu}\xi_j^{\mu}}=\me^{-\frac{|j-i|}{L}},
		\label{eq:correlation}
	\end{equation}
	where $L=L(\beta)$ is the correlation length at inverse temperature $\beta$ and satisfies 
	\begin{align}
		L=-\frac{1}{\ln\qua*{\tanh\ton*{\beta}}}.
		\label{eq:correlation_length}
	\end{align}
	
	Having made these considerations, the procedure for assessing the dynamical stability of patterns is the same depicted in Sec.~\ref{sec:s2n} and so we have to compute the mean value and the variance of $\xi^{1}_ih_i\ton*{\vec{\xi}^{1}}$. By using \eqref{eq:field} we have that the mean value satisfies\footnote{Note that we already made the approximation $\frac{N-1}{N}=1$ since we are interested in the large $N$ limit.}
	\begin{align*}
		\mean*{h_1\xi_1^1}&=1+\frac{1}{N}\sum_{\mu=2}^K\sum_{j>1}^N\mean*{\xi_1^{\mu}\xi_j^{\mu}\xi_1^{1}\xi_j^{1}}=\\
		&=1+\frac{1}{N}\sum_{\mu=2}^K\sum_{j>1}^N\mean*{\xi_1^{\mu}\xi_j^{\mu}}\mean*{\xi_1^{1}\xi_j^{1}},
	\end{align*}
	where we used the fact that different patterns are uncorrelated. Exploiting the expression of the two point correlations \eqref{eq:correlation} we get
	\begin{align*}
		\mean*{h_1\xi_1^1}&=1+\frac{1}{N}\sum_{\mu=2}^K\sum_{j>1}^N\exp\qua*{-2\frac{j-1}{L}}=\\
		&=1+\frac{K-1}{N}\gra*{-1+\sum_{x=0}^{N-1}\qua*{\exp\ton*{-\frac{2}{L}}}^x},
	\end{align*}
	where we defined $x=j-1$. In the limit of large $N$, this yields
	\begin{equation}
		\mean*{h_1\xi_1^1}=1+\frac{K-1}{N}\qua*{\frac{1}{1-\me^{-2/L}}-1}=1+\frac{K-1}{N}\frac{1}{\me^{2/L}-1}.
		\label{eq:expectation_local_field}
	\end{equation}
	The quantity $\mean*{\ton*{h_1\xi_1^1}^2}$ can obtained by very similar passages, but since the computation is quite lengthy we only report the final result; the interested reader can find the detailed computation in Appendix~\ref{app:variance}. It holds
	\begin{align}
		\mean*{\ton*{h_1\xi_i^1}^2}=&\frac{K-1}{N}+4\frac{K-1}{N}\frac{1}{\me^{2/L}-1}+\nonumber\\
		&+\frac{(K-1)(K-2)}{N^2}\frac{1}{\me^{2/L}-1}+\nonumber\\
		&+2\frac{(K-1)(K-2)}{N^2}\frac{1}{\ton*{\me^{2/L}-1}^2}+1.
		\label{eq:expectation_local_field_squared}
	\end{align}
	By putting this expression and \eqref{eq:expectation_local_field} into the condition for dynamical stability given by \eqref{eq:stability_patterns_simpler} we conclude that patterns can be retrieved up to a critical correlation length $L_c$ satisfying 
	\begin{align*}
		C-\alpha+\frac{\qua*{2(C-1)-4}\alpha}{\me^{2/L_c}-1}+\frac{(C-1)\alpha^2}{\ton*{\me^{2/L_c}-1}^2}-\frac{\alpha^2}{\me^{2/L_c}-1}=0.
	\end{align*}
	We recall that here $\alpha=K/N$ is the load of the Hopfield neural network.
	Solving for $L_c$ this yields
	\[
		L_c=\frac{2}{\log\qua*{\frac{\alpha^2+\alpha\sqrt{4+\alpha^2-4C}+2C-2\alpha C}{2(C-\alpha)}}}
	\]
	and so the inverse temperature $\beta_c$ of the one dimensional Ising model above which retrieval breaks due to spatial correlations is
	\[
		\beta_c=\tanh^{-1}\qua*{\sqrt{\frac{2(C-\alpha)}{\alpha^2+\alpha\sqrt{4+\alpha^2-4C}+2C-2\alpha C}}}.
	\]

	Now, if we set $\beta_c=0$ we return to uncorrelated Rademacher patterns and so to the original \gls{hnn} for which the maximal load is $\alpha_c\approx0.14$. As a consequence we can fix $C$ by imposing $T_c(\alpha_c, C)=\infty$, which gives $C=\alpha_c\approx0.14$. In conclusion the $\beta_c$ above which Hopfield neural network can no more handle the spatial correlation of patterns is
	\begin{equation}
		\beta_c(\alpha)=\tanh^{-1}\qua*{\sqrt{\frac{2(\alpha_c-\alpha)}{\alpha^2+\alpha\sqrt{4+\alpha^2-4\alpha_c}+2\alpha_c-2\alpha \alpha_c}}}.
		\label{eq:beta_critical_hopfield}
	\end{equation}
	This last expression allows us to draw a phase diagram in the variables $\beta, \alpha$ dividing a fuzzy phase from a retrieval phase. This is shown in Fig.~\ref{fig:H2-T_ic_vs_alphafull}, where we also plotted the results of numerical simulations confirming the validity of \eqref{eq:beta_critical_hopfield}. As we mentioned in the introduction, when data show internal correlation they carry a lower amount of information. One would thus expect them to be easier to store and so the capacity of the neural network to increase, while the results we obtained go in the opposite direction. This apparently paradoxical behavior can be explained noticing that also the coupling matrix, which is defined starting from the patterns themselves, is less complex than in the uncorrelated case and thus if on one side data are easier to memorize, on the other the neural network has de facto less resources to handle them. 

\section{Dense Associative Memories with correlated patterns}
    \label{sec:dam}
    We now move to the case of Dense Associative Memories storing correlated patterns generated by the 1D Ising model at temperature $\beta$. The procedure is analogous to that followed in Sec.~\ref{sec:s2n}, but as done in Sec.~\ref{sec:hopfield} we now have to consider correlations among the different components of the patterns. We only consider an Hamiltonian with interactions involving an odd number of neurons since in this way both computations and numerical simulations are easier to perform. As shown in the Appendix~\ref{app:dam} in the case under consideration it can be proven that
    \[
        \mean*{h_1^{(p)}\xi_1^1}=1
    \]
    and
    \begin{align*}
		\mean*{\ton*{h_1^{(p)}\xi_1^1}^2}=1+\sum_{n=0}^{p-1}&\left\{\frac{K-1}{N^{2p-2-n}}\frac{\qua*{(p-1)!}^2}{n!\qua*{(p-n-1)!}^2}\right.\\
		&\left.\qua*{\frac{2N}{\me^{2/L}-1}}^{p-n-1}\right\}.
	\end{align*}
	From \eqref{eq:stability_patterns_simpler} we thus obtain that retrieval is possible up to a maximal correlation length $L_c$ satisfying
	\[
		\frac{K-1}{N^{p-1}}\sum_{n=0}^{p-1}\gra*{\frac{\qua*{(p-1)!}^2}{n!\qua*{(p-n-1)!}^2}\qua*{\frac{2}{\me^{2/L_c}-1}}^{p-n-1}}=C.
	\]
	In the high load case $K=\alpha N^{p-1}$ this yields
	\[
		\sum_{n=0}^{p-1}\gra*{\frac{\qua*{(p-1)!}^2}{n!\qua*{(p-n-1)!}^2}\qua*{\frac{2}{\me^{2/L_c}-1}}^{p-n-1}}=\frac{C}{\alpha}.
	\]
	Analogously to the case $p=2$ the constant $C$ can be determined requiring $L_c$ to be zero when $\alpha=\alpha_c^{(p)}$. Here by $\alpha_c^{(p)}$ we denote the critical load of the $p$-spin DAM with uncorrelated patterns above which retrieval breaks. Note that sometimes the load of DAM is defined as $\alpha=K/N$ as in the Hopfield case, but we prefer to set $\alpha=K/N^{p-1}$ in order to make it possible to make a comparison for different values of $p$. For $L_c=0$, the patterns becomes spatially uncorrelated and so only the term with $n=p-1$ should be considered, as a consequence we get 
	\[
		(p-1)!=\frac{C}{\alpha_c^{(p)}} \ \to \ C=(p-1)!\alpha_c^{(p)}.
	\]
	In conclusion our implicit expression for $L_c$ is 
	\[
		\sum_{n=0}^{p-1}\gra*{\frac{\qua*{(p-1)!}}{n!\qua*{(p-n-1)!}^2}\qua*{\frac{2}{\me^{2/L_c}-1}}^{p-n-1}}=\frac{\alpha_c^{(p)}}{\alpha}.
	\]
	For $p=3$ this last equation can be solved analytically, indeed we have
	\[
		\frac{2}{\ton*{\me^{2/L_c}+1}^2}+\frac{4}{\me^{2/L_c}+1}+1=\frac{\alpha_c^{(3)}}{\alpha},
	\]
	whose solution is 
	\[
		L_c^{(3)}=\frac{2}{\log\ton*{\frac{\alpha_c^{(3)}+\sqrt{2\ton*{\alpha_c^{(3)}\alpha+\alpha^2}}+\alpha}{\alpha_c^{(3)}-\alpha}}}.
	\]
	The critical inverse temperature $\beta_c^{(3)}$ is therefore 
	\begin{equation}
		\beta_c^{(3)}=\tanh^{-1}\qua*{\sqrt{\frac{\alpha_c^{(3)}-\alpha}{\alpha_c^{(3)}+\sqrt{2\ton*{\alpha_c^{(3)}\alpha+\alpha^2}}+\alpha}}}
		\label{eq:beta_c_p3}
	\end{equation}
	\begin{figure}
	    \centering
	    \includegraphics[width=\columnwidth]{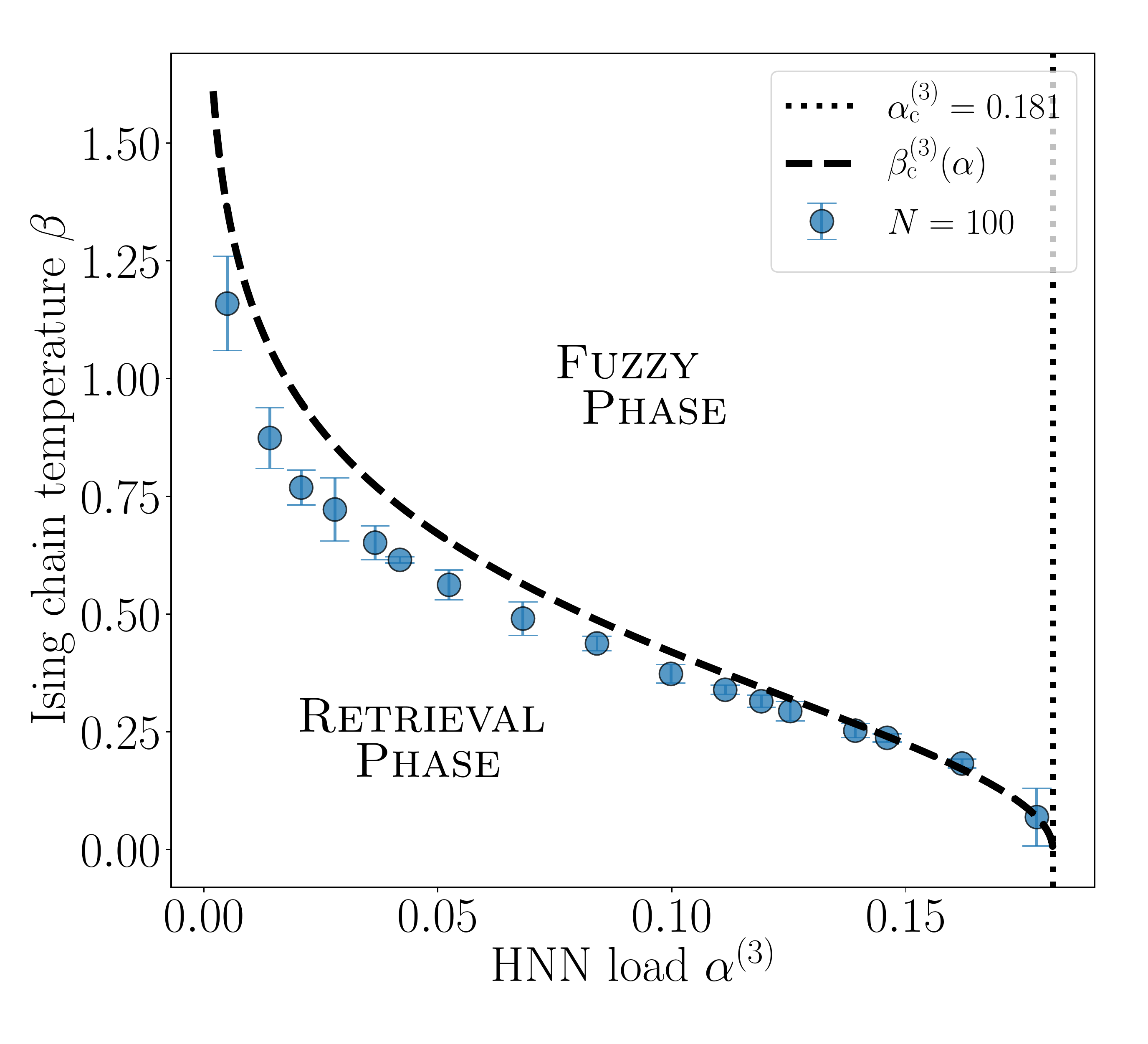}
	    \caption{\textbf{Numerical simulations for a $p=3$ DAM endowed with spatially correlated patters.} We simulated a DAM with $p=3$-body interaction, $N=100$ neurons and varying load $\alpha$ under the dynamics \eqref{eq:evolution} and using as starting state one of the patterns $\vec{\sigma}=\vec{\xi^{\mu}}$ generated using a linear Ising chain at temperature $\beta$. We then collected the Mattis magnetization $m_{\mu}$ for each possible starting pattern, we repeated the process $M=3$ times and we computed the average magnetization $\mean*{m^{\mu}}$. For any $\alpha$ we computed the critical $\beta_c^{(3)}$ above which retrieval breaks as the value for which the average magnetization is equal to Abbot's magnetization $m=0.807$ \cite{abbott1987storage}, which is the equivalent of Sompolinsky magnetization for the case $p=3$. Since for $p=3$ the critical load $\alpha_c^{(3)}$ is unknown, we left it as a free parameter and we fitted the theoretical prediction \eqref{eq:beta_c_p3} to simulated data (dashed black line). Uncertainties have been computed as the semi-difference between the last point above Abbot's threshold and the one just below. }
	    \label{fig:H3-T_ic_vs_alphafull}
	\end{figure}
	Again we have an expression allowing us to sketch a phase diagram in the variables $\beta, \alpha$ dividing, as for the Hopfield case, a fuzzy phase from a retrieval phase. This is shown in Fig.~\ref{fig:H3-T_ic_vs_alphafull}, where we also plotted the results of numerical simulations. Note that while for $p=2$ an analytical estimate of $\alpha_c$ as been computed, for the case $p=3$ only approximated values are available. As a consequence we decided to leave $\alpha_c^{(3)}$ as a free parameter, which we then estimated by fitting \eqref{eq:beta_c_p3} to the simulated data. Results are shown in Fig.~\ref{fig:H3-T_ic_vs_alphafull}a and from this procedure we obtain $\alpha_c^{(3)}\approx 0.18$. This value is larger than analytical results obtained under the replica symmetry assumption and confirms the expected increase of $\alpha_c^{(3)}=$ when also replica symmetry breaking is taken into account \cite{abbott1987storage}. 
\section{Conclusions}
    \label{sec:conclusions}
    Artificial neural networks are nowadays used in uncountable applications, but despite the great interest they gathered we are still lacking a general theory describing how and why they work. Understanding the mechanism behind their ability in handling complex problems is thus a central issue and many scientists are working in this direction. Hopfield neural network is one of the few examples where analytical results can be obtained and great effort has been devoted to draw phase diagrams for this particular system. Indeed, despite it being very simple if compared to modern deep neural networks, the Hopfield model still show a very rich phenomenology characterized by different phase transitions. Moreover, this neural network has also been shown to be equivalent to Restricted Boltzmann Machines, which are used in a number of real life applications. However, in order to make the model feasible to be analyzed with the tools of statistical mechanics, one often has to make strong assumption on the nature of the information to be stored. Indeed, one generally considers the case of i.i.d. random data, which is quite an unrealistic scenario since in most real life situations patterns are correlated one with the other or they are structured, thus showing spatial correlations. 
    
    In the present paper we considered how the HNN behaves when feed with spatially correlated data, that we generated from a linear Ising chain with inverse temperature $\beta$. These synthetic patterns can be seen as prototypes of textual data and their spatial correlation can be tuned by changing the temperature of the Ising chain. By using a signal to noise approach, we obtain an analytical expression for the critical inverse temperature $\beta_c(\alpha)$ above which correlation becomes too strong and retrieval breaks. This allows to draw a phase diagram in the variables $\alpha$, $\beta$ where a fuzzy phase and a retrieval region can be identified. Remarkably, we observe both numerically and analytically that the introduction of spatial correlations diminishes the critical load of the Hopfield network, while one would naively expect the opposite to occur. These results are then extended to Dense Associative Memories, namely generalized Hopfield networks with p-body interactions. In particular we consider the case of odd p and we derive again an expression for the critical inverse temperature $\beta^{(p)}_c(\alpha)$. As in the $p=2$ case corresponding to the HNN, we observe the critical load to diminish when spatial correlations are increased, a result which is also confirmed by numerical simulations performed for a DAM with $p=3$.
    
    Spatial mono-dimensional correlations, typical of textual data, are only an example of the possible structure real patterns can show; as also mentioned in the introduction, another straightforward example is given by two dimensional correlations in images. As the 1d Ising chain configurations can be used as prototypes of textual data, in the very same way synthetic spatially correlated images could be generated by means of a two dimensional Ising model. Extending the analysis here performed to such a case would be a very interesting problem to investigate in a future publication, also considering the non-trivial n-body correlation emerging in the 2d Ising model.
    
    \acknowledgements
        We are grateful to Elena Agliari, Adriano Barra and Francesco Alemanno for their useful comments and discussion. We acknowledge the project “Social and Economic Complexity” of Enrico Fermi Research Center.

	
%
    \appendix

\section{Detailed computations for \texorpdfstring{\gls{hnn}}{HNN}}
    \label{app:variance}
    Here we show how to compute the quantity $\mean*{\ton*{h_1\xi_i^1}^2}$ feed with one dimensional Ising configurations. By virtue of \eqref{eq:expression_squared_field} we have
	\[
		\ton*{h_1\xi_1^1}^2=\underbrace{2h_1\xi_1^1-1}_{A}+\underbrace{\frac{1}{N^2}\sum_{j,k=2}^N\sum_{\mu, \rho=2}^K\xi_1^{\mu}\xi_j^{\mu}\xi_j^{1}\xi_1^{\rho}\xi_k^{\rho}\xi_k^{1}}_{B}
	\]
	The expectation of the first term can be easily computed using \eqref{eq:expectation_local_field}
	\begin{equation}
		\mean*{A}=2\frac{K-1}{N}\frac{1}{\me^{2/L}-1}+1.
		\label{eq:mean_A}
	\end{equation}
	For what concerns term $B$ we first have to get rid of self-interactions. In particular we rewrite it as
	\begin{align*}
		B=\frac{1}{N^2}\sum_{\mu, \rho =2}^K\Bigg[&\sum_{j\neq1}^N\xi_1^{\mu}\xi_j^{\mu}\xi_1^{\rho}\xi_k^{\rho}+\\
		&+\sum_{j>1}^N\sum_{k\neq 1,j}^N\xi_1^{\mu}\xi_j^{\mu}\xi_j^{1}\xi_1^{\rho}\xi_k^{\rho}\xi_k^{1}\Bigg]
	\end{align*}
	and going on we get
	\begin{align*}
		B=\frac{1}{N^2}\Bigg\{&\sum_{\mu\neq1}^K\Bigg[\sum_{j\neq1}^N+\sum_{j>1}^N\sum_{k\neq 1,j}^N\xi_j^{\mu}\xi_j^{1}\xi_k^{\mu}\xi_k^{1}\Bigg]+\\
		+&\sum_{\mu\neq1}^K\sum_{\rho\neq1, \mu}^K\Bigg[\sum_{j\neq1}^N\xi_1^{\mu}\xi_j^{\mu}\xi_1^{\rho}\xi_k^{\rho}+\sum_{j>1}^N\sum_{k\neq 1,j}^N\xi_1^{\mu}\xi_j^{\mu}\xi_j^{1}\xi_1^{\rho}\xi_k^{\rho}\xi_k^{1}\Bigg]\Bigg\}.
	\end{align*}
	We thus have four terms 
	\begin{align}
		B=&\underbrace{\frac{(K-1)(N-1)}{N^2}}_{\text{I}}+\underbrace{\frac{1}{N^2}\sum_{\mu\neq1}^K\sum_{j>1}^N\sum_{k\neq 1,j}^N\xi_j^{\mu}\xi_j^{1}\xi_k^{\mu}\xi_k^{1}}_{\text{II}}+\nonumber\\
		+&\underbrace{\frac{1}{N^2}\sum_{\mu\neq1}^K\sum_{\rho\neq1, \mu}^K\sum_{j\neq1}^N\xi_1^{\mu}\xi_j^{\mu}\xi_1^{\rho}\xi_j^{\rho}}_{\text{III}}+\nonumber\\
		&+\underbrace{\frac{1}{N^2}\sum_{\mu\neq1}^K\sum_{\rho\neq1, \mu}^K\sum_{j>1}^N\sum_{k\neq 1,j}^N\xi_1^{\mu}\xi_j^{\mu}\xi_j^{1}\xi_1^{\rho}\xi_k^{\rho}\xi_k^{1}}_{\text{IV}}.
		\label{eq:B}
	\end{align}
	The average of term I is trivial, for what concerns term II we have
	\begin{align*}
		\mean*{\text{II}}&=\frac{1}{N^2}\sum_{\mu\neq1}^K\sum_{j>1}^N\sum_{k\neq 1,j}^N\mean*{\xi_j^{\mu}\xi_k^{\mu}}\mean*{\xi_j^{1}\xi_k^{1}}=\\
		&=\frac{K-1}{N^2}\sum_{j>1}^N\sum_{k\neq 1,j}^N\exp\ton*{-2\frac{\abs{k-j}}{L}}=\\
		&=2\frac{K-1}{N^2}\sum_{j>1}^N\sum_{k>j}^N\exp\ton*{-2\frac{j-k}{L}}.
	\end{align*}
	Defining $x=k-j$ we get
	\begin{align*}
		\mean*{\text{II}}&=2\frac{K-1}{N^2}\sum_{j=2}^N\qua*{-1+\sum_{x=0}^{N-j}\ton*{\me^{-2/L}}^x}\\
		&=2\frac{K-1}{N^2}\sum_{j=2}^N\frac{\me^{2(j-N)/L}-1}{1-\me^{2/L}}=\\
		&=2\frac{K-1}{N^2}\frac{1}{1-\me^{2/L}}\qua*{-(N-2)+\sum_{y=0}^{N-2}\ton*{\me^{-2/L}}^y},
	\end{align*}
	where we defined $y=N-j$. For $N\to\infty$ the sum, being convergent, is a negligible correction and therefore we can write 
	\begin{equation}
		\mean*{\text{II}}=2\frac{K-1}{N}\frac{1}{\me^{2/L}-1}.
		\label{eq:II}
	\end{equation}
	Going on to term III we have 
	\begin{align*}
		\mean*{III}=&\frac{1}{N^2}\sum_{\mu\neq1}^K\sum_{\rho\neq1, \mu}^K\sum_{j\neq1}^N\mean*{\xi_1^{\mu}\xi_j^{\mu}}\mean*{\xi_1^{\rho}\xi_j^{\rho}}\\
		&=\frac{(K-1)(K-2)}{N^2}\sum_{j=2}^N\exp\qua*{-2\frac{j-1}{L}},
	\end{align*}
	that is, for large $N$
	\begin{equation}
		\mean*{III}=\frac{(K-1)(K-2)}{N^2}\frac{1}{\me^{2/L}-1}.
		\label{eq:III}
	\end{equation}
	Finally we have term IV
	\begin{align*}
		\mean*{\text{IV}}&=\frac{1}{N^2}\sum_{\mu\neq1}^K\sum_{\rho\neq1, \mu}^K\sum_{j>1}^N\sum_{k\neq 1,j}^N\mean*{\xi_1^{\mu}\xi_j^{\mu}}\mean*{\xi_1^{\rho}\xi_k^{\rho}}\mean*{\xi_j^{1}\xi_k^{1}}\\
		&=2\frac{(K-1)(K-2)}{N^2}\sum_{j=2}^N\sum_{k>j}^N\me^{-(j-1)/L}\me^{-(k-1)/L}\me^{-(k-j)/L}.
	\end{align*}
	Proceeding as done before and considering $N\gg1$ one finds
	\begin{equation}
		\mean*{IV}=2\frac{(K-1)(K-2)}{N^2}\frac{1}{\ton*{\me^{2/L}-1}^2}
		\label{eq:IV}
	\end{equation}
	Exploiting Eqs.~\eqref{eq:B}, \eqref{eq:II}, \eqref{eq:III} and \eqref{eq:IV}, the expectation of $B$ is then
	\begin{align}
		\mean*{B}=&\frac{K-1}{N}+2\frac{K-1}{N}\frac{1}{\me^{2/L}-1}+\frac{(K-1)(K-2)}{N^2}\frac{1}{\me^{2/L}-1}+\nonumber\\
		&+2\frac{(K-1)(K-2)}{N^2}\frac{1}{\ton*{\me^{2/L}-1}^2}
		\label{eq:mean_B}
	\end{align}
	and combining this expression with \eqref{eq:mean_A} we obtain for the expectation of $\ton*{h_1\xi_i^1}^2$
	\begin{align}
		\mean*{\ton*{h_1\xi_i^1}^2}=&\frac{K-1}{N}+4\frac{K-1}{N}\frac{1}{\me^{2/L}-1}+\nonumber\\
		&+\frac{(K-1)(K-2)}{N^2}\frac{1}{\me^{2/L}-1}+\nonumber\\
		&+2\frac{(K-1)(K-2)}{N^2}\frac{1}{\ton*{\me^{2/L}-1}^2}+1.
		\label{eq:mean_field_squared}
	\end{align}
	
\section{Detailed computations for the Dense Associative Memories}
    \label{app:dam}
    Here we show how to compute the quantities $\mean*{h_1^{(p)}\xi_1^1}$ and $\mean*{\ton*{h_1\xi_1^1}^2}$ for a DAM with odd bodies interaction. Using \eqref{eq:local_field_pbody} we can write the former as
	\begin{align*}
		\mean*{h_1\xi_1^1}&=\frac{1}{N^{p-1}}\sum_{\mu=1}^K\sum_{i_2\ldots i_p}^N\mean*{\xi_1^{\mu}\xi_1^1\prod_{k=i_2\ldots i_p}\xi_k^{\mu}\xi_k^1}=\\
		&=1+\frac{1}{N^{p-1}}\sum_{\mu=2}^K\sum_{i_2\ldots i_p}^N\mean*{\xi_1^{\mu}\xi_1^1\prod_{k=i_2\ldots i_p}\xi_k^{\mu}\xi_k^1}.
	\end{align*}
	Since different patterns are uncorrelated, we can rewrite the expectation as 
	\[
		\mean*{\xi_1^{\mu}\xi_1^1\prod_{k=i_2\ldots i_p}\xi_k^{\mu}\xi_k^1}=\mean*{\xi_1^{\mu}\prod_{k=i_2\ldots i_p}\xi_k^{\mu}}\mean*{\xi_1^{1}\prod_{k=i_2\ldots i_p}\xi_k^{1}}=0,
	\]	
	where the last equality derives from the fact that the expectation of an odd number of spins is null for the 1D Ising model. Concluding we have 
	\begin{equation}
		\mean*{h_1\xi_1^1}=1.
		\label{eq:expectation_field_p_odd}
	\end{equation}
	
	We can now turn to the expectation of the local field squared so to compute the variance. Analogously to \eqref{eq:expression_squared_field} we have
	\begin{equation}
		\ton*{h_1\xi_1^1}^2=\underbrace{2h_1\xi_1^1-1}_{A}+\underbrace{\qua*{\frac{1}{N^{p-1}}\sum_{\mu=2}^K\sum_{i_2\ldots i_p}^N\xi_1^{\mu}\xi_1^1\prod_{k=i_2\ldots i_p}\xi_k^{\mu}\xi_k^1}^2}_{B}
		\label{eq:A+B_expectation_squared_filed_p_body_odd}
	\end{equation}
	The expectation of the first term is trivial and satisfies $\mean*{A}=1$, so let us focus on the last term, its expectation is 
	\[
		\mean*{B}=\frac{1}{N^{2p-2}}\sum_{\mu,\rho=2}^K\sum_{i_2\ldots i_p}^N\sum_{j_2\ldots j_p}^N\mean*{\xi_1^{\mu}\xi_1^{\rho}\prod_{k=i_2\ldots i_p}\xi_k^{\mu}\xi_k^1\prod_{l=j_2\ldots j_p}\xi_l^{\rho}\xi_l^1},
	\]
	that is 
	\begin{align*}
		\mean*{B}=\frac{1}{N^{2p-2}}\sum_{\mu,\rho=2}^K\sum_{i_2\ldots i_p}^N\sum_{j_2\ldots j_p}^N&\mean*{\prod_{k=i_2\ldots i_p}\prod_{l=j_2\ldots j_p}\xi_k^1\xi_l^1}\\
		&\mean*{\xi_1^{\mu}\xi_1^{\rho}\prod_{k=i_2\ldots i_p}\prod_{l=j_2\ldots j_p}\xi_k^{\mu}\xi_l^{\rho}}.
	\end{align*}
	If $\mu\neq\rho$, the second expectation can be factorized in two expectations each containing an odd number of terms which thus give a null contribution. We can then set $\mu=\rho$ obtaining
	\begin{align*}
		\mean*{B}=\frac{1}{N^{2p-2}}\sum_{\mu=2}^K\sum_{i_2\ldots i_p}^N\sum_{j_2\ldots j_p}^N&\mean*{\prod_{k=i_2\ldots i_p}\prod_{l=j_2\ldots j_p}\xi_k^1\xi_l^1}\\
		&\mean*{\prod_{k=i_2\ldots i_p}\prod_{l=j_2\ldots j_p}\xi_k^{\mu}\xi_l^{\mu}},
	\end{align*}
	which yields
	\begin{equation}
		\mean*{B}=\frac{K-1}{N^{2p-2}}\sum_{i_2\ldots i_p}^N\sum_{j_2\ldots j_p}^N\mean*{\prod_{k=i_2\ldots i_p}\prod_{l=j_2\ldots j_p}\xi_k^1\xi_l^1}^2
		\label{eq:term_B_expectation_field_squared_pbody}
	\end{equation}
	Let us focus on the expectation, by writing it in full we have
	\[
		\mean*{\prod_{k=i_2\ldots i_p}\prod_{l=j_2\ldots j_p}\xi_k^1\xi_l^1}=\mean*{\xi_{i_2}^1\xi_{i_3}^1\ldots\xi_{i_p}^1\xi_{j_2}^1\xi_{j_3}^1\ldots\xi_{j_p}^1}.
	\]
	Here we have to pay attention to all terms in which two or more of the $\xi$s are equal. Since no self-interaction is present in the \gls{hnn}, this can occur only if $i_a=j_b$ for some $a, b$. As a consequence we have to count in how many ways we can form $n$ pairs between the $\xi_i$ and the $\xi_j$. It is easy to see that such a number $M(n, p)$ is given by 
	\[
		M(n, p)=\frac{\qua*{(p-1)!}^2}{n!\qua*{(p-n-1)!}^2},
	\]
	because the first pair can be chosen in $\qua*{(p-1)!}^2$ ways, the second in $\qua*{(p-2)!}^2$ and so on, but we have also to divide by $n!$ in order to not to give any relevance to the order by which the different pairs are chosen. Exploiting this result we can rewrite \eqref{eq:term_B_expectation_field_squared_pbody} as 
	\begin{widetext}
		\begin{align}
			\mean*{B}&=\frac{K-1}{N^{2p-2}}\sum_{i_2\ldots i_p}^N\sum_{j_2\ldots j_p}^N\sum_{n=0}^{p-1}\qua*{\frac{\qua*{(p-1)!}^2}{n!\qua*{(p-n-1)!}^2}\mean*{\prod_{k=i_2\ldots i_{p-n}}\prod_{l=j_2\ldots j_{p-n}}\xi_k^1\xi_l^1}^2\prod_{m=p-n+1}^{p}\delta_{i_mj_m}}=\nonumber\\
			&=\sum_{n=0}^{p-1}\qua*{\frac{K-1}{N^{2p-2-n}}\frac{\qua*{(p-1)!}^2}{n!\qua*{(p-n-1)!}^2}\sum_{i_2\ldots i_{p-n}}^N\sum_{j_2\ldots j_{p-n}}^N\mean*{\prod_{k=i_2\ldots i_{p-n}}\prod_{l=j_2\ldots j_{p-n}}\xi_k^1\xi_l^1}^2}=\nonumber\\
			&=\sum_{n=0}^{p-1}\gra*{\frac{K-1}{N^{2p-2-n}}\frac{\qua*{(p-1)!}^2}{n!\qua*{(p-n-1)!}^2}\qua*{\sum_{k=2}^N\sum_{l=2, l\neq k}^N\mean*{\xi_k^1\xi_l^1}^2}^{p-n-1}}.
		\end{align}
	\end{widetext}
	Note that in writing this expression we are using the convention that when there is not an explicit kronecker delta making two indices equal they are different. Moreover we exploited the fact that in the one dimensional Ising model the correlation between an even number of spins can be factorized in two-bodies correlations. We can finally rewrite the expectation of $B$ recurring to \eqref{eq:II}, this gives
	\[
		\mean*{B}=\sum_{n=0}^{p-1}\gra*{\frac{K-1}{N^{2p-2-n}}\frac{\qua*{(p-1)!}^2}{n!\qua*{(p-n-1)!}^2}\qua*{\frac{2N}{\me^{2/L}-1}}^{p-n-1}}
	\]
	From this result, going back to \eqref{eq:A+B_expectation_squared_filed_p_body_odd}, we obtain
	\begin{align*}
		\mean*{\ton*{h_1\xi_1^1}^2}=1+\sum_{n=0}^{p-1}&\left\{\frac{K-1}{N^{2p-2-n}}\frac{\qua*{(p-1)!}^2}{n!\qua*{(p-n-1)!}^2}\right.\\
		&\left.\qua*{\frac{2N}{\me^{2/L}-1}}^{p-n-1}\right\}
	\end{align*}
\end{document}